\documentclass[prb,10pt,twocolumn,superscriptaddress,footnoteinbib]{revtex4}
\usepackage{amsmath}
\usepackage{latexsym}
\usepackage{amssymb}
\usepackage{graphics,epstopdf}
\usepackage[colorlinks=true, citecolor=blue, urlcolor=blue ]{hyperref}
\usepackage{epsf,graphics,graphicx}

\date{\today}

\begin{document}

\title{Thermal signature of helical molecule: Beyond nearest-neighbor electron hopping}

\author{Suparna Sarkar}

\email{physics.suparna@gmail.com}

\affiliation{Theoretical Sciences Unit, Jawaharlal Nehru Centre for Advanced Scientific Research, Bangalore-560 064, India}

\author{Santanu K. Maiti}

\email{santanu.maiti@isical.ac.in}

\affiliation{Physics and Applied Mathematics Unit, Indian Statistical Institute, 203 Barrackpore Trunk Road, Kolkata-700 108, India}

\author{David Laroze}

\email{dlarozen@uta.cl}

\affiliation{Instituto de Alta Investigaci\'{o}n, Universidad de Tarapac\'{a}, Casilla 7D, Arica, Chile}

\begin{abstract}

We investigate, for the first time, the thermal signature of a single-stranded helical molecule that is described beyond usual 
nearest-neighbor electron hopping, by analyzing electronic specific heat. Depending on the hopping of electrons, two different 
kinds of helical systems are considered. In one case the hopping is confined within a few neighboring lattice sites which is 
referred to as short-range hopping helix, while in the other case, electrons can hop in all possible sites making the system 
a long-range hopping one. These two helices accurately emulate the structures of single-stranded DNA and protein molecules, 
respectively. Each helix geometry is exposed to a transverse electric field applied perpendicular to the helix axis. Due to 
this field, the system transforms into a correlated disordered one, resembling the well-known Aubry-Andr\'e-Harper (AAH) model.
The interplay among the helicity, higher-order hopping, and the electric field has significant impact on thermal response.
Our comprehensive theoretical analysis reveals that, under low-temperature conditions, the short-range hopping helix exhibits 
greater sensitivity to temperature compared to the long-range hopping helix system. Conversely, the scenario reverses in the 
high-temperature limit. The thermal response of the helices can be modified selectively by means of the electric field, and 
the difference between the specific heats of the two helices gradually decreases with increasing the field strength. The 
molecular handedness, whether left-handed or right-handed, on the other hand does not have any appreciable effect on the 
thermal signature. In addition, we also explore a significant application of electronic specific heat (ESH). If the helix contains 
a point defect, then by comparing the results of perfect and defective helices, one can estimate the location of the defect, 
which might be useful in diagnosing bad cells and different diseases. Finally, we discuss the results of ESH by 
considering the spin degree of freedom and in the context of real biological helical systems.       

\end{abstract}

\maketitle

\section{Introduction}

During the past few years, helical molecules have been actively used for designing different efficient nanoscale charge and 
spin-based electronic devices~\cite{dev1,dev2,dev3}. Starting from spin-specific electron transmission, current rectification, 
magneto-resistive behavior, and many other operations are performed with great 
efficiency~\cite{hel1,hel2,hel3,hel4,hel5,hel6,hel7,hel8,sp1,sp2,sp3,sp4,sp5,recti,sp6,sp7,gmr,thermo}. All these functionalities 
are directly involved with the unique and diverse characteristic features of helical molecular systems. Depending on 
the physical parameters, essentially two different kinds of helical molecules are defined. In one case, electrons are able to hop 
from one atomic site to its few neighboring sites, making the system a short-range hopping helix~\cite{hel5}. Whereas, in another 
case, electrons can hop in all possible atomic sites yielding a long-range hopping helix~\cite{hel5,hel6}. Structurally, the most common 
and realistic examples of long-range hopping (LRH) helices are single-stranded protein molecules, while short-range hopping (SRH) helices 
are typically exemplified by DNA molecules~\cite{hel6}. Due to the existence of higher-order electron hopping, these systems exhibit 
several unusual phenomena compared to the traditional nearest-neighbor hopping model. 

{\em While significant progress has been achieved in investigating electron and spin-dependent transport phenomena 
across various real biological and custom-designed `helical' systems~\cite{bio1,bio2,bio3}, the thermal signature of helical systems 
remains unexplored to the best of our knowledge. This gap in research undoubtedly requires proper attention and exploration.} 
In a study conducted in $2015$, Kundu and Karmakar have investigated~\cite{sh7} the thermal response of a double-stranded DNA (ds-DNA) 
molecule by analyzing the behavior of ESH within a tight-binding framework. Their research focused on four distinct sequences of ds-DNA, 
considering variations in disorderness. The central objective was to explore the role of the backbone structure, in conjunction 
with environmental interactions. The DNA molecules were characterized using only the nearest-neighbor hopping model, with no 
consideration given to helicity effects. Several additional studies are also available~\cite{sh1,sh2,sh3,sh4,sh5,sh6,sh8,sh9,sh10,sh11,kk1}, 
with each one limited to the nearest-neighbor hopping of electrons. {\em However, none of these studies have endeavored to investigate the 
thermal response of a system that incorporates higher-order electron hopping along with helicity}.

\begin{figure}[ht]
\vskip 0.5cm
\noindent
{\centering \resizebox*{4.5cm}{9cm}{\includegraphics{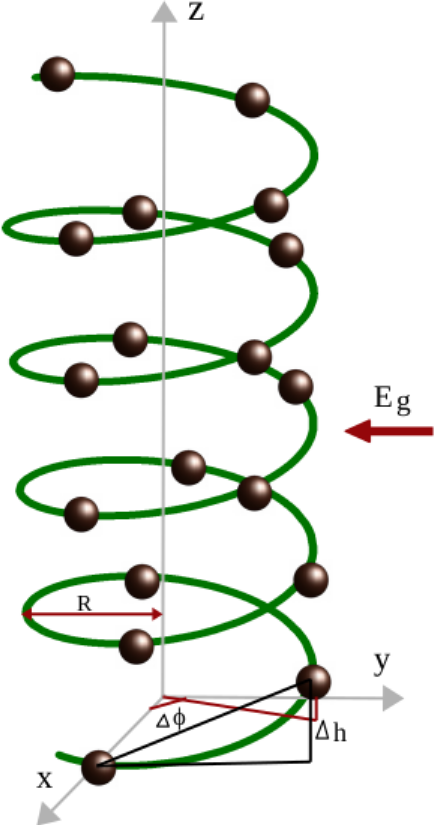}}\par}
\caption{(Color online). Schematic view of a right-handed single-stranded helical molecule, where $R$ denotes the radius, $\Delta \phi$ 
represents the twisting angle and $\Delta h$ corresponds to the stacking distance between the neighboring lattice sites. The helix is 
subjected to a perpendicular electric field of strength $E_g$. In the presence of this field, the system becomes a correlated disordered
one.}
\label{model}
\end{figure}

In this current study, we explore the thermal signature of a physical system that extends beyond the conventional nearest-neighbor 
electron hopping, incorporating the effect of helicity. Two different kinds of single-stranded helices, SRH and LRH, are taken into 
account depending on the electron hopping range. For a possible tuning of ESH and in order to inspect the role of disorder, an electric
field is applied perpendicular to the helix axis (see Fig.~\ref{model}). Because of the helicity, site energies of the system get 
modulated~\cite{efl,lhrh} in a cosine form, following the well-known Aubry-Andr\'{e}-Harper model~\cite{flores,ab1,ab2,ab3,ab4,ab5}. 
This is a special class of disorder (correlated) that exhibits several non-trivial signatures in the context of electronic
localization~\cite{loc1,loc2,loc3,loc4}. Unlike a one-dimensional random (uncorrelated) disordered lattice where all the eigenstates 
are localized irrespective of the disorder strength for the nearest-neighbor hopping model~\cite{dis1,dis2}, an AAH system goes to 
the localized phase from the conducting 
phase beyond a finite disorder strength~\cite{ab1,ab2,ab3}. More interesting phenomena are observed when the system is described beyond 
the nearest-neighbor hopping~\cite{sdsarma}. The modulation in site energies directly influences the energy eigenspectrum, and thus, 
the electronic specific heat. It will be very interesting and important as well to check the interplay between the helicity, 
higher-order electron hopping, and the electric field on ESH. Here it is relevant to point out that, in their work Kundu and Karmakar
have studied~\cite{sh7} the response of ESH in presence of `substitutional disorder' introducing disorder in the backbone sites 
attached to the ds-DNA molecule. Both quasi-periodic and random forms were taken into account. But in our chosen physical system,
substitutional disorder is absent, instead the disorder arises from the helicity of the geometry in the presence of an external 
electric field. Furthermore, the disorder follows the AAH form, imparting distinctive features, particularly in situations where the 
system involves higher-order electron hopping~\cite{sdsarma}.

The helical molecular system is described using a tight-binding (TB) framework, which always provides a straightforward level of 
description~\cite{hel5,hel7,sp7}. 
Diagonalizing the Hamiltonian matrix for the helix, we determine the average energy. Subsequently, by computing the derivative of 
the average energy with respect to temperature, we obtain the specific heat~\cite{sh3,sh6,sh7}. Through our 
comprehensive numerical analysis, we find that the sensitivity of an SRH helix to temperature surpasses that of the LRH 
helix in the low-temperature range. Conversely, as temperature rises, the response becomes inverted. Despite the presence of an electric 
field, the distinctive characteristics of both temperature regimes persist, albeit with a gradual reduction in the disparity of specific 
heats between the SRH and LRH helices as the field strength increases. It is noteworthy that the electric field can selectively modify 
the thermal response, but the handedness exhibits no significant impact. Our work critically examines all these aspects, providing 
compelling arguments for our findings. In the end, we focus on a significant application of ESH. Notably, the concept of ESH has been 
identified as a potential tool for diagnosing various neurodegenerative diseases such as Parkinson, Alzheimer, Creutzfeldt-Jakob, 
and others~\cite{shd}. In our study, we introduce a point defect at an arbitrary lattice site within our helices and endeavor to 
identify the defect's location by comparing the outcomes of perfect and defective helices. This approach holds promise in the diagnosis 
of defective cells and various diseases, contributing to the potential application of this protocol in medical contexts. 

It is worth noting that most of the results in this work are analyzed for custom-designed helical geometries for 
spinless electrons. However, for the sake of completeness, we also discuss results considering (i) real biological systems with random 
sequences of DNA bases, adenine (A), thymine (T), guanine (G), and cytosine (C), and (ii) the spin degree of freedom, in the 
appropriate subsections.

The novel and crucial aspects of our ESH study include: (i) exploring the interplay among helicity, higher-order hopping, 
and electric field, (ii) investigating the combined effects of temperature and the range of electron hopping, and (iii) determining 
defective lattice sites.

The remaining portion of the work is structured as follows. The helical systems and the theoretical framework used for the calculations 
are given in Sec. II. Section III presents and meticulously examines all numerical results. Finally, Section IV encapsulates a summary 
of the findings.

\section{Physical system, tight-binding Hamiltonian and theoretical formulation}

\subsection{Helix geometry and tight-binding Hamiltonian}

The physical system, whose thermal signature is the subject of discussion in this work, is illustrated schematically in Fig.~\ref{model}.
It is a right-handed helix. Two crucial parameters that fundamentally characterize a helical structure are the twisting angle 
$\Delta \phi$ (indicating the degree of geometric twist) and the stacking distance $\Delta h$ between two consecutive lattice
sites~\cite{hel5,sp7}. 
Depending on the value of $\Delta h$, two types of helices are typically considered. When the distance between neighboring sites is 
small (i.e., $\Delta h$ is low), electrons can easily traverse from one site to all other possible sites in the geometry, defining the 
system as a long-range hopping helix. Conversely, in the case where $\Delta h$ is high, indicating large atom separation, electron 
hopping is confined to a few neighboring atomic sites, characterizing the system as a short-range hopping helix. Although LRH and SRH 
helices may appear structurally identical, their behaviors differ significantly. In this study, one of our objectives is to investigate
the distinct roles of these two forms of electron hopping in influencing the thermal signature.

An electric field is applied perpendicular to the helix axis. There are two primary motivations for applying such a field. Firstly, 
in the presence of this field, the system transforms into a correlated (non-random) disordered state~\cite{sp7,efl} from an ordered 
one. Consequently, 
the impact of disorder can be systematically examined. Exploring the interplay between disorder and the range of electron hopping 
on thermal response could yield intriguing observations. Secondly, the aim is to investigate how the externally tuned parameter can 
modify the thermal response. These aspects have not been explored in the existing literature.

A tight-binding framework is presented to elucidate the helical system. The general form of the TB Hamiltonian for a $N$-site helix, 
whether it is short-range or long-range, is expressed as~\cite{hel5} 
\begin{eqnarray}
H & = & \sum_i \mbox{{$\epsilon$}}_i \mbox{{$c$}}_i^{\dagger} \mbox{{$c$}}_i \nonumber \\
& + & 
\sum_{i=1}^{N-1}\sum_{j=1}^{N-i} \left(\mbox{{$t$}}_j \mbox{{$c$}}_i^{\dagger} \mbox{{$c$}}_{i+j} + 
\mbox{{$t$}}_j^* \mbox{{$c$}}_{i+j}^{\dagger} \mbox{{$c$}}_{i}\right)
\label{equ1}
\end{eqnarray} 
where $\epsilon_i$ denotes the site energy of an electron at $i$th lattice site, and $c_i^{\dagger}$, $c_i$ represent 
the fermionic creation and annihilation operators, respectively. $t_j$ is the hopping strength between the sites $i$ and ($i+j$).
$t_j^*$ is the complex conjugate of $t_j$, and in our case, it is identical to $t_j$, since $t_j$ is real.
In terms of the nearest-neighbor hopping strength $t_1$, $t_j$ can be expressed as~\cite{hel7,qfs}
\begin{eqnarray}
t_j=t_1 \exp[-(l_j-l_1)/l_c]
\label{equ2}
\end{eqnarray}
where $l_c$ is the decay constant and $l_j$ is the distance of separation between the sites $i$ and ($i+j$). $l_1$ measures the
nearest-neighbor distance. The length $l_j$ is expressed in terms of the structural parameters of the helix as~\cite{hel7,qfs}
\begin{eqnarray}
l_j=\sqrt{[2R\sin(j\Delta\phi/2)]^2+(j\Delta h)^2}
\label{equ3}
\end{eqnarray}
where $R$ represents the radius (see Fig.~\ref{model}).
It is important to note that in our chosen tailored helical systems, we assume that the hopping strength depends only on 
the distance between two sites, without considering site energy dependence. However, in real molecular systems like DNA and proteins, 
electron hopping depends on the specific sites as well as the distance between them~\cite{dh1,dh2,dh3,dh4}. When we will be discussing 
thermal response of real biological molecules in later part of this work, these effects will be taken into account.

Now, when an electric field is applied perpendicular to the helix axis, the site energies undergo modulation in a specific form 
following the relation~\cite{efl}
\begin{equation}
\epsilon_{i}^\prime = \epsilon_i + e V_g \cos\left(i\Delta \phi-\beta\right).
\label{equ4}
\end{equation}
Here, $e$ is the electronic charge and $V_g$ is the gate voltage that is responsible for producing the field. The gate voltage and the
electric field are related by the equation $V_g=E_g R$~\cite{efl}. The parameter $\beta$, refereed to as the phase factor, describes 
the field direction with respect to the positive $X$ axis. At this point, it is noteworthy to highlight that the aforementioned 
representation of site energies bears a resemblance to the widely recognized Aubry-Andr\'e-Harper model~\cite{ab1,ab2,ab3}, wherein 
$eV_g$ measures the strength of cosine modulation. Consequently, our system exhibits a substantial equivalence to the AAH model, 
incorporating higher-order hopping of electrons.

\subsection{Theoretical formulation}

The electronic specific heat is determined by calculating the first order derivative of average energy of the system with respect to 
temperature $T$. At constant volume it is defined as~\cite{sh3,sh6,sh7} 
\begin{equation}
C_v = \frac{\partial \bar E}{\partial T} 
\label{equ5}
\end{equation}
where $\bar E$ is the average energy, and it is expressed as~\cite{sh3,sh6,sh7}
\begin{equation}
\bar E = \sum_{i=1}^{N} \left(E_i-\mu\right) f(E_i).
\label{equ6}
\end{equation}
The average energy is obtained by taking the weighted average of the energy levels, where the weights correspond to the 
probabilities of finding electrons in these energy levels having eigenenergies $E_i$. The eigenvalues are determined by numerically
diagonalizing the TB Hamiltonian matrix of the helix system. In the above expression (Eq.~\ref{equ6}), $\mu$ is the electro-chemical 
potential and $f(E_i)$ is the Fermi-Dirac distribution function. At temperature $T$, 
$f(E_i)$ is given by
\begin{equation}
f(E_i) = \left(1+e^\frac{E_i-\mu}{K_B T}\right)^{-1}
\label{equ7}
\end{equation}
where $K_B$ is the Boltzmann constant. Substituting $f(E_i)$ in Eq.~\ref{equ6}, and doing some straightforward algebra we get the 
form of the electronic specific heat, $C_v$, as
\begin{equation}
C_v = \frac{1}{K_BT^2}\sum_{i=1}^{N} \frac{\left(E_i-\mu\right)^2 e^\frac{E_i-\mu}{K_BT}}{\left(1+e^\frac{E_i-\mu}{K_B T}\right)^2}.
\label{equ8}
\end{equation}

Since $C_v$ is intricately linked to the electronic density of states (DOS) of the system, we additionally calculate 
it to provide a lucid depiction of our findings. The density of states, denoted as $\rho(E)$, is acquired through the 
relation~\cite{gf1}
\begin{equation}
\rho(E)=-\frac{1}{\pi} \mbox{Im}[\mbox{Tr}[G(E)]]
\label{dos}
\end{equation}
where $G(E)$ represents the Green's function of the helix system. The Green's function is defined as~\cite{gf1,gf2}
\begin{equation}
G(E)=\frac{1}{E+i \eta-H}
\label{grf}
\end{equation}
where $\eta \rightarrow 0$.

\section{Numerical results and discussion}

In this section, we present our numerical results encompassing the characteristic features of the electronic specific heat for both
custom-designed and realistic LRH and SRH helices, in scenarios with and without an applied electric field. Throughout 
the calculations, energies are measured in electron-volt (eV), temperature is measured in Kelvin (K), and the Boltzmann constant 
$K_B$ is set at $8.6173303 \times 10^{-5}\,$eV/K. For the custom-designed helices, unless explicitly specified otherwise, we choose the 
site energy $\epsilon_i=0$ and the nearest-neighbor hopping strength $t_1=0.02\,$eV. The reason for selecting $t_1$ in 
this range is to make it comparable to real helical biological systems. For the helical systems with A, T, G, and C sites, 
the site energies and hopping strengths are taken from the standard data set available in the literature~\cite{sh6,zil}, 
and the values are specified in the relevant sub-sections of our discussion. The site energies of any helix, be it 
synthetic or real, 
become non-uniform following the prescription stated earlier, once the electric field is applied. Unless mentioned otherwise, the
results are for right-handed helices with $N=20$ in the absence of spin degree of freedom (DOF). For completeness, the results of ESH
considering spin DOF are also discussed in appropriate parts. In each case, the electro-chemical potential $\mu$ is determined 
self-consistently by fixing the total number of electrons, $N_e$, in the system. The results are worked out in the half-filled band cases, 
and unless specifically dictated, they are for synthetic helical systems. 

As previously noted, the prevalent and practical instances of SRH and LRH helices are exemplified by single-stranded DNA 
\begin{table}[ht]
\caption{Physical parameters of the right-handed SRH and LRH helices.}
\vskip 0.15cm
\begin{tabular}{|c|c|c|c|c|}
\hline \hline
Helix & $R$ ($\mbox\AA$) & $\Delta h$ ($\mbox\AA$) & $\Delta\phi$ (rad) & $l_c$ ($\mbox\AA$)  \\ \hline
SRH & $7$ & $3.5$ & $\frac{\pi}{5}$ & $0.9$  \\ \hline
LRH & $2.5$ & $1.5$ & $\frac{5 \pi}{9}$ & $0.9$  \\ \hline \hline
\end{tabular}
\label{tab1}
\end{table}   
\begin{table}[ht]
\caption{Few neighboring distances for the SRH and LRH helices, considering the physical parameters given in Table~\ref{tab1}.}
\vskip 0.15cm
\begin{tabular}{|c|c|c|c|c|c|}
\hline \hline
Helix & $l_1$  & $l_2$ & $l_3$ & $l_4$ & $l_5$  \\ \hline 
SRH & $5.502$  & $10.675$ & $15.242$ & $19.033$ & $22.023$  \\ \hline 
LRH & $4.113$  & $5.766$ & $5.148$ & $6.239$ & $8.850$  \\ \hline \hline 
\end{tabular} 
\label{tab2}
\end{table}
and protein molecules, respectively~\cite{hel5}. In our computations, we define the physical parameters for these helices based on 
the standard dataset accessible in the existing literature~\cite{rps}. They are described in Table~\ref{tab1}. 
For a concise depiction of how the electron hopping is influenced by the physical parameters associated with the helices, in
Table~\ref{tab2} we display the distances between a few neighboring sites in our chosen SRH and LRH helices, as the hopping is 
directly connected to the distance of separation between the sites. The data reveals that for the SRH helix, the neighboring distances
exhibit a rapid increase with the site index. Consequently, electron hopping is confined to a few neighboring atomic sites.
In contrast, the LRH helix presents a markedly different scenario, with distances remaining relatively comparable even as the site index 
reaches large values. This characteristic allows electron hopping to occur across all possible sites, introducing non-trivial features 
compared to conventional lattices possessing only nearest-neighbor hopping.
Though the exponential form in the expression of the hopping integral is considered, this parameter setting clearly shows 
that for the LRH helix, electrons can hop to distant sites with moderate strength. This type of parameter setting has been extensively
considered in the literature for other studies. Thus, long-range hopping persists effectively.

We now proceed to present and analyze our results which include (i) the interplay among the higher order electron hopping, transverse 
electric field and the helicity, (ii) combined effects of temperature and the range of electron hopping,  (iii) determination of 
defective lattice sites, (iv) role of spin degree of freedom, and (v) response of real helical systems. 
The results are organized into distinct sub-sections.

\subsection{In absence of external electric field}

\begin{figure}[ht]
\vskip 0.5cm
\noindent
{\centering\resizebox*{6.5cm}{4.5cm}{\includegraphics{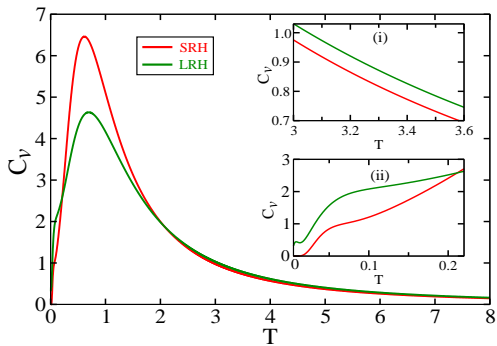}}}
\caption{(Color online). Specific heat versus temperature for the SRH and LRH helices in the zero field case. The insets show the 
characteristics in the high- and low-temperature limits, for a specific temperature range, to have better viewing of the two colored 
curves.}
\label{cvt}
\end{figure}

Let us start our discussion by referring to Fig.~\ref{cvt}, which illustrates the variation of 
electronic specific heat as a function
of temperature both for the SRH and LRH helices, under the condition of no external electric field, i.e., $V_g=0$. Several noteworthy
features emerge from this figure that are outlined as follows. In the low-temperature limit, the specific heat exhibits a rapid increase 
with temperature. Upon reaching a maximum, it then gradually decreases and ultimately approaches to zero in the high-temperature limit. 
This behavior is a shared characteristic of both the SRH and LRH cases, as evident from the curves presented in Fig.~\ref{cvt}.
An intriguing observation is that, in the case of a helix with short-range electron hopping, the specific heat attains a significantly 
larger magnitude compared to another helical system under low-temperature conditions. As temperature increases, the disparity between 
the curves gradually diminishes. Notably, beyond a specific temperature range, the specific heat of the LRH helix takes precedence over 
the SRH counterpart. This distinctive behavior is discerned from the results presented in inset (i) of Fig.~\ref{cvt}.
\begin{figure}[ht]
\vskip 0.5cm
\noindent
{\centering\resizebox*{7cm}{7.5cm}{\includegraphics{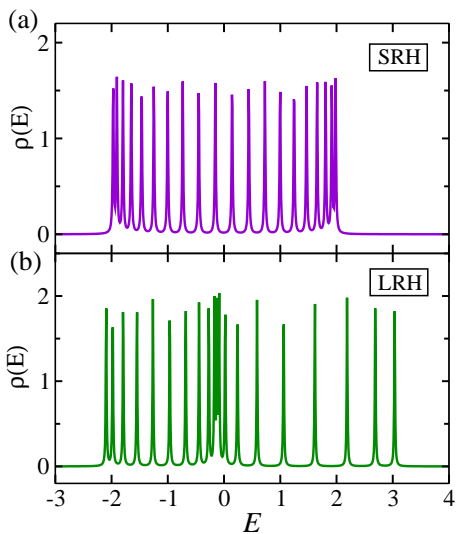}}}
\caption{(Color online). DOS as a function of energy for the two helices in the zero field case.}
\label{dosvg0}
\end{figure}
The aforementioned observations can be elucidated by referring to the DOS spectra depicted in Fig.~\ref{dosvg0}, as the specific heat is
intricately linked to the electronic density of states. At zero temperature, all electronic energy levels up to the Fermi energy (for
finite temperatures we refer chemical potential, instead of Fermi energy) are completely filled, while the remaining energy levels 
remain unoccupied. Upon introducing temperature, a subset of neighboring energy levels within the $k_BT$ range around the Fermi energy 
begins to contribute, thereby augmenting the average energy ($\bar{E}$) in comparison to the zero-temperature scenario. As the 
temperature rises, a greater number of energy levels become accessible, leading to an increase in the average energy and, consequently, 
a higher electronic specific heat. The rate of enhancement for $\bar{E}$ in the SRH helix surpasses that of the LRH helix. 
This is attributed to the specific nature of the DOS profile of the SRH helix, as illustrated in Fig.~\ref{dosvg0}(a).
The eigenstates are denser towards the edges of the allowed energy region, while in the inner energy regions, the gaps between the
DOS peaks are quite comparable. This kind of behavior has already been previously reported~\cite{dh3,ds1,ds2}. The DOS profile of the
SRH helix clearly suggests that a greater number of energy levels become accessible within a specific range. In contrast, for the LRH 
system, the energy levels exhibit substantial gaps in the higher energy region, as depicted in Fig.~\ref{dosvg0}(b), consistent with the
previous studies~\cite{hel5,sp7}. Therefore, a smaller number of eigenenergies is available. 

As the temperature rises, the likelihood of occupation in higher energy levels gradually increases, causing all energy levels to 
contribute to the average energy $\bar{E}$. 
Hence, achieving a further enhancement of $\bar{E}$ with temperature becomes less feasible. This results in a reduction of $C_v$. 
Eventually, when the temperature is reasonably high, the average energy almost stabilizes, and under that condition, the specific 
heat approaches to insignificantly small values. Since the LRH helix exhibits energy levels at higher energies compared to the SRH 
helix (followed by comparing the DOS spectra given in Fig.~\ref{dosvg0}), in the high-temperature limit, the rate of change of $\bar{E}$ 
with temperature for the LRH helix surpasses that of the SRH helix. Hence, a higher $C_v$ is attained. For a clearer representation, 
we refer to inset (i) of Fig.~\ref{cvt}.

In Fig.~\ref{cvt}, an additional intriguing feature becomes apparent at the extremely low-temperature limit, specifically when the 
system's temperature approaches nearly zero. The specific heat of the LRH helix surpasses that of the SRH helix within a very 
limited temperature range (see inset (ii) of Fig.~\ref{cvt}). This is solely due to the highly packed energy levels of the LRH helix 
around $\mu$ within the $K_BT$ range (Fig.~\ref{dosvg0}(b)), whereas for the SRH helix no such behavior is available 
(Fig.~\ref{dosvg0}(a)).
 
\subsection{In presence of external electric field}

We now turn our attention to examining the impact of an external electric field on thermal response.  
\begin{figure}[ht]
\vskip 0.5cm
\noindent
{\centering \resizebox*{8.5cm}{4.2cm}{\includegraphics{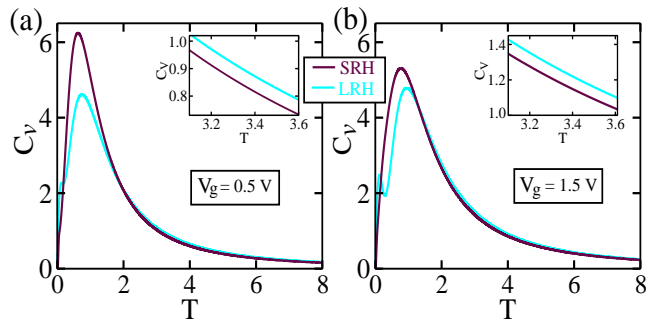}}\par}
\caption{(Color online). Electronic specific heat as a function of temperature for the SRH and LRH helices in presence of external 
electric field where (a) and (b) correspond to $V_g=0.02\,$V and $0.04\,$V, respectively. The phase factor $\beta$ is set at zero. The 
inset in each sub-figure corresponds to the same meaning as mentioned in the inset (i) of Fig.~\ref{cvt}.}
\label{cvtvg}
\end{figure}
In Fig.~\ref{cvtvg}, we illustrate 
the variation in electronic specific heat with temperature for two distinct helices under the influence of an external electric field, 
i.e., $V_g\ne 0$. Interestingly, the overall $C_v$-$T$ spectra exhibit a striking resemblance to those observed in the absence of a 
field. Specifically, there is a sharp increase in specific heat at low temperatures, reaching a maximum before gradually diminishing 
and ultimately vanishing at high temperatures. Notably, the LRH helix consistently demonstrates a higher value of $C_v$ compared to the 
SRH helix, both in the limit of zero temperature and in the higher temperature range, mirroring the behavior observed in the field-free
case. However, a meticulous analysis reveals that, in the presence of the electric field, the disparity between the peak values of the 
$C_v$-$T$ curves associated with the SRH and LRH helices diminishes. This effect becomes more pronounced as the strength of the external 
field increases. To comprehend this behavior, we delve into the following analysis.

\begin{figure}[ht]
\vskip 0.5cm
\noindent
{\centering\resizebox*{8cm}{6cm}{\includegraphics{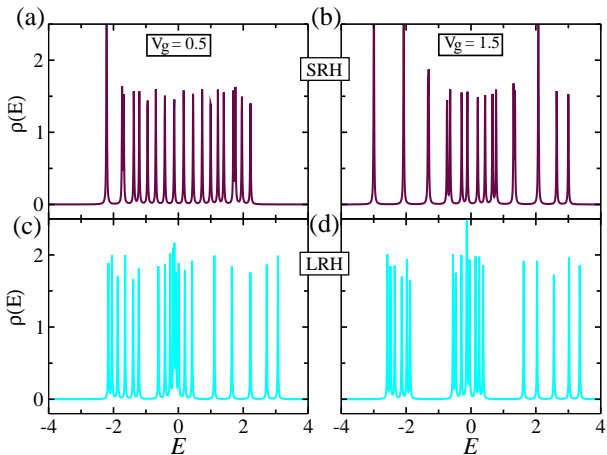}}}
\caption{(Color online). DOS as a function of energy for the SRH and LRH helices at the same gate voltages as taken in Fig.~\ref{cvtvg},
where the first and second columns correspond to $V_g=0.02\,$V and $0.04\,$V, respectively. Here we set $\beta=0$.}
\label{dosvg}
\end{figure}

The central mechanism becomes evident when examining the density of states. In Fig.~\ref{dosvg}, we present the DOS spectra for the two 
helices, corresponding to the gate voltages illustrated in Fig.~\ref{cvtvg}.
The impact of the electric field is truly intriguing. Upon the application of the electric field, the site energies undergo modulation in
a cosine form as described by the relation given in Eq.~\ref{equ4}. This cosine modulation transforms the perfect helical
system into a correlated disordered one, analogous to the well-known AAH model~\cite{ab1,ab2,ab3}. Due to this distinctive modulation, 
the energy spectrum becomes fragmented and exhibits large gaps, resembling the organization of energy levels into three sub-bands. 
A clearer visualization of 
this phenomenon emerges for a larger system size with a moderate field strength, and similar energy spectra have been extensively 
explored in a series of papers focusing on AAH systems~\cite{ab1,ab2,ab3}. For a substantial $V_g$ (keeping in mind the value of $t_1$, 
as mentioned earlier), the impact of correlated 
disorder takes precedence over electron hopping. Consequently, both the SRH and LRH helices exhibit quite comparable DOS spectra, as
illustrated in the right column of Fig.~\ref{dosvg}, and hence their specific heats do not vary significantly (Fig.~\ref{cvtvg}(b)).
In contrast, when $V_g$ is lower, the DOS spectra for the two helices show substantial differences (left column of Fig.~\ref{dosvg}),
and accordingly, a notable contrast between the specific heats is evident, as shown in Fig.~\ref{cvtvg}(a).
 
Due to the pronounced impact of the electric field on electronic specific heat, it is essential to investigate how $C_v$ undergoes
\begin{figure}[ht]
\vskip 0.5cm
\noindent
{\centering\resizebox*{7cm}{4.5cm}{\includegraphics{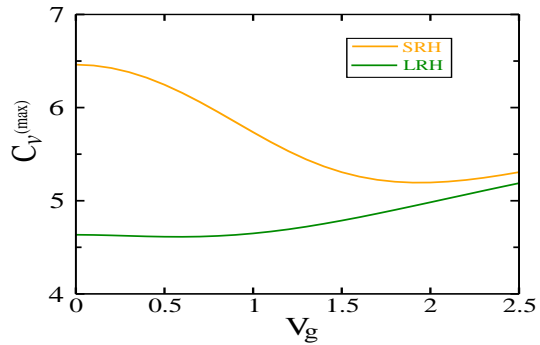}}}
\caption{(Color online). $C_v$(max)-$V_g$ characteristics for the SRH and LRH helices. Here $\beta=0$.}
\label{cvmvg}
\end{figure}
modifications with a continuous variation of the field strength, or alternatively, by varying the correlated disorder strength. 
To accomplish this, we calculate the maximum of $C_v$ denoted as $C_v$(max) and depict it as a function of $V_g$. For each value 
of $V_g$, we determine $C_v$ across a broad temperature range and subsequently identify the maximum to obtain $C_v$(max). The results 
for both the helices are presented in Fig.~\ref{cvmvg}, with the orange and green lines corresponding to the SRH and LRH helices, 
respectively. In accordance with the aforementioned analysis, it is noteworthy that the disparity between the specific heats is most 
prominent at $V_g=0$ and gradually diminishes with an increase in the field strength. For sufficiently large $V_g$, the specific heats 
converge and become nearly indistinguishable. Since the nearest-neighbor hopping strength $t_1$ is within the meV range, the upper 
limit of $V_g$ used in this figure (Fig.~\ref{cvmvg}) (viz, $V_g=0.06\,$V) is considered sufficiently large. From the curves, we 
can infer that the electric field plays a pivotal role in modulating the thermal response.

To ensure a comprehensive analysis, we now explore the dependencies of electronic specific heat on various physical parameters 
inherent to the system.

\vskip 0.25cm
\noindent
$\blacksquare$ {\bf Role of $\beta$}: To examine the influence of the electric 
\begin{figure}[ht]
\vskip 0.5cm
\noindent
{\centering\resizebox*{8cm}{4cm}{\includegraphics{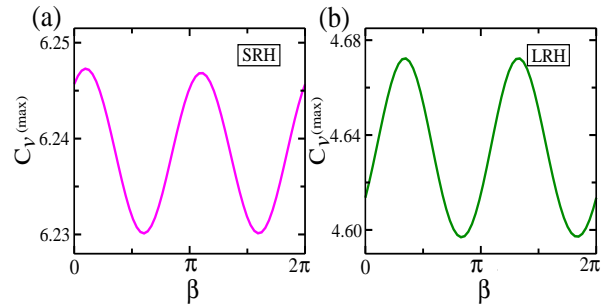}}}
\caption{(Color online). Dependence of $C_v$(max) with $\beta$ for the two helical systems when $V_g=0.02\,$V.}
\label{cvmbeta}
\end{figure}
field direction on the thermal response, in 
Fig.~\ref{cvmbeta} we plot the variation of $C_v$(max) with $\beta$ for both the SRH and LRH helices, with the gate voltage 
fixed at $0.02\,$V. The parameter $\beta$, linked to the field direction, is incorporated into the site energy expression through
Eq.~\ref{equ4}. The site energies undergo modifications with changes in $\beta$, leading to alterations in the spectrum of DOS 
and subsequently affecting the specific heat. A comparison between SRH and LRH systems reveals that, while the variation is not 
significantly different in both cases, the LRH helix exhibits a slightly higher response than the SRH helix.

\vskip 0.25cm
\noindent
$\blacksquare$ {\bf Effect of geometrical conformation}: The recognized consensus affirms that the stacking distance 
$\Delta h$ between adjacent lattice sites plays a crucial role in defining the extent of electron hopping.
\begin{figure}[ht]
\vskip 0.5cm
\noindent
{\centering \resizebox*{8cm}{4.2cm}{\includegraphics{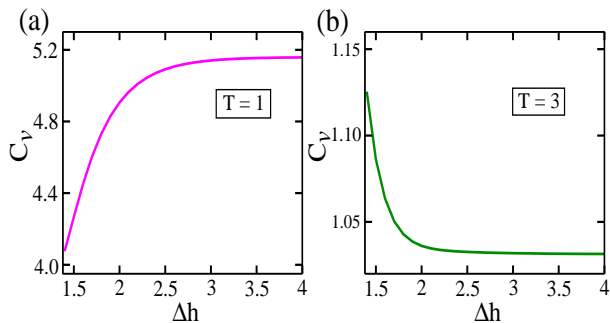}}\par}
\caption{(Color online). Dependence of $C_v$ with the stacking distance $\Delta h$ for the two helices at two different temperatures.
The other physical parameters are $V_g=0.04\,$V and $\beta=0$.}
\label{cvdh}
\end{figure}
When $\Delta h$ is too small, electrons have the ability to hop across all available sites. Conversely, when $\Delta h$ is relatively 
large, electron hopping becomes confined to a limited number of neighboring sites. Since the electron hopping range significantly
influences the energy band spectrum, thereby impacting the density of states, here we explore the influence of $\Delta h$ (viz,
the geometrical conformation) on electronic specific heat. The findings are illustrated in Fig.~\ref{cvdh}. Two temperature scenarios,
low and high, are considered. Across both temperature ranges, we vary $\Delta h$ from $1.5\AA$ (the chosen value for our LRH helix, 
as detailed in Table~\ref{tab1}) to a larger value, up to $4\AA$. Consistent with our earlier assertions, we observe that at low 
temperatures, the electronic specific heat increases as $\Delta h$ rises, i.e., when the system becomes more short-ranged 
(Fig.~\ref{cvdh}(a)). Conversely, at high temperatures, a contrasting trend emerges, characterized by a reduction in $C_v$ as 
$\Delta h$ increases (Fig.~\ref{cvdh}(b)). These trends can be directly attributed to the alterations in the DOS resulting from 
changes in the electron hopping range.

\vskip 0.25cm
\noindent
$\blacksquare$ {\bf Effect of chirality}: At this juncture, it is crucial to investigate the sensitivity of the thermal response to
\begin{figure}[ht]
\vskip 0.5cm
\noindent
{\centering \resizebox*{8.5cm}{4.2cm}{\includegraphics{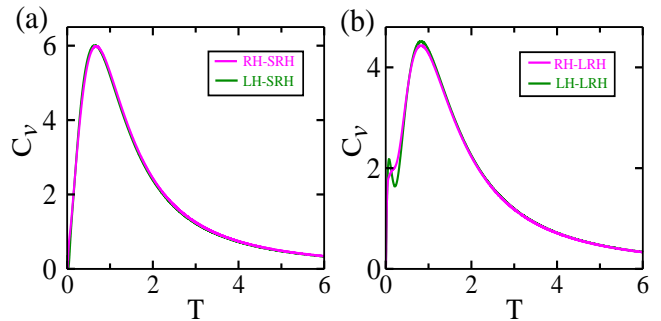}}\par}
\caption{(Color online). Role of chirality on the electronic specific heat. In each of the spectra, the results of right-handed (RH) and
left-handed (LH) helices are shown, where (a) and (b) are associated with the SRH and LRH systems, respectively. The other parameters 
are $V_g=0.02\,$V, $\beta=\pi/4$, $N=21$ (for SRH helix), and $N=19$ (for LRH helix).}
\label{chirality}
\end{figure}
\begin{figure}[ht]
\vskip 0.5cm
\noindent
{\centering\resizebox*{7cm}{7cm}{\includegraphics{doschirality.eps}}}
\caption{(Color online). DOS spectra for the right-handed (dark blue color) and left-handed (cyan color) helices, 
employing the identical set of parameter values as depicted in Fig.~\ref{chirality}. Panels (a) and (b) correspond to the SRH and 
LRH cases, respectively.}
\label{dch}
\end{figure}
the chirality of the helical system. The findings thus far have been deduced for right-handed helices. However, in the presence of 
an electric field, there is a potential for a modified density of states based on the handedness of the system, as the site energies 
undergo alterations with the shift from right-handed (RH) to left-handed (LH) chirality. To achieve a left-handed helix, it is necessary 
to substitute $\Delta \phi$ with $-\Delta \phi$~\cite{lhrh}, while keeping all other factors constant. Consequently, site energies differ 
for RH and LH cases, in accordance with the relationship defined in Eq.~\ref{equ4}, leading us to anticipate a change in thermal 
signature. It is worth noting that for a seamless transition between handedness, the system size must be appropriately chosen to 
ensure the helix completes full turns. Depending on the structural parameters specified in our SRH and LRH helices (as detailed 
in Table~\ref{tab1}), we opt for $N=21$ for the SRH case (equivalent to two turns) and $N=19$ for the LRH helix (equivalent to five turns). 
In Fig.~\ref{chirality}, we present the results of ESH, considering both right-handed and left-handed helices at a typical gate voltage. 
It is observed that, in each hopping case, the specific heats exhibit almost comparable values for both the right- and left-handed systems.
Upon changing the chirality, the DOS spectrum does not undergo significant modifications, as clearly demonstrated by comparing the DOS 
spectra for left- and right-handed helices in Fig.~\ref{dch}. Consequently, nearly identical curves are obtained for the two different
handedness configurations.
 
\subsection{Usefulness of electronic specific heat}

In this sub-section, we explore a specific application of the ESH, focusing on its utility in identifying defective lattice sites 
within a system by analyzing the thermal response. Our objective is to investigate the possibility of detecting any anomalies 
in the lattice structure. To achieve this, we introduce a single defect at an arbitrary site while keeping all other sites of the 
helix identical.
\begin{figure}[ht]
\vskip 0.5cm
\noindent
{\centering \resizebox*{8.5cm}{4.2cm}{\includegraphics{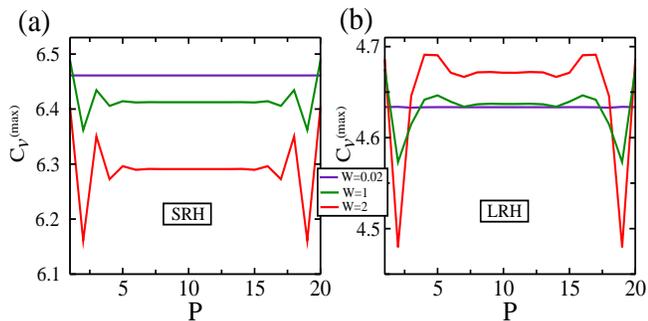}}\par}
\caption{(Color online). Variation of $C_v$(max) with the position of the defective site in the SRH and LRH helices. The results are 
shown for three distinct values of $W$. Here we consider the zero-field case. The system size $N$ is fixed at $20$.}
\label{disorder}
\end{figure}
The strength of the site energy of the defected site is denoted by the parameter $W$ ($W=0$ corresponds to the perfect site, as the 
site energies of all other sites are set at zero). We present three distinct cases based on varying values of $W$: one with a weak 
defect ($W=0.0002$), and the other two with moderate and substantial defects. For each case, we calculate the maximum of the ESH, 
$C_v$(max), and illustrate its variation as a function of the position of the defective site, represented by the parameter $P$ (refer 
to Fig.~\ref{disorder}). It is noteworthy that we deliberately choose a very small value for $W$ in one case ($<<t_1$) to mimic a 
scenario where the system closely resembles the perfect one. This allows us to make meaningful comparisons with finite $W$ and gain 
insights into the behavior of the system under different defect strengths.
Several noteworthy features are discerned, outlined as follows. For $W=0.0002$, $C_v$(max) remains nearly constant for both SRH and LRH 
helices (violet lines in Fig.~\ref{disorder}), aligning with expectations. However, a pronounced shift in $C_v$(max) occurs at $W=0.02$ 
when the defective site is positioned near one edge of the helix (green lines), with this effect becoming more conspicuous for larger $W$ 
(red lines). Although the $C_v$(max)-$P$ curves exhibit similarities between the two helices, a discernible alteration in ESH with 
varying $P$ is observed for the helix with higher-order hopping integrals. By comparing the outcomes of perfect and defective helices, 
a clear distinction can be made, allowing for the identification of the defective helix and an approximate determination of its location.
Particularly, when the impurity site is situated near either of the two edges, a substantial variation is observed, presenting a 
valuable indicator for precise detection.
      
\subsection{Electronic specific heat in presence spin degeneracy}

To inspect whether the inclusion of spin degree of freedom has any important impact, in Fig.~\ref{cvsp} we present the
\begin{figure}[ht]
\vskip 0.5cm
\noindent
{\centering \resizebox*{7.5cm}{5cm}{\includegraphics{cv_T_vg0_spin.eps}}\par}
\caption{(Color online). Electronic specific heat, considering spin degree of freedom, as a function of temperature for 
the SRH and LRH helices in the field-free condition ($V_g=0$), for the half-filled band case ($N=20$, $N_e=20$). Other parameters and
the meaning of the insets are same as in Fig.~\ref{cvt}.}
\label{cvsp}
\end{figure}
variation of electronic specific heat as a function of temperature for the two different helices considering spin degeneracy.
The results are shown in the field-free situation, for the half-filled band condition. As our chosen $N$ is $20$, the number of electrons
$N_e$ is fixed to $20$. Here each energy level can accommodate two electrons. The electro-chemical potential $\mu$ is calculated 
self-consistently, as in all other figures. Apart from spin 
degeneracy, all other conditions remain unchanged from Fig.~\ref{cvt}. Comparing the results between Figs.~\ref{cvt} and \ref{cvsp}, it is
clear that the latter figure reflects only a change by a factor of $2$, with no other differences. This is easy to follow, as the average 
energy doubles in the presence of spin degeneracy compared to the spin-free case. Thus, we can emphasize that spin degeneracy does not 
introduce any new characteristic features to specific heat, apart from changing its magnitude.

\subsection{Electronic specific heat in real helical systems}

The studied results of synthetic helices create curiosity about the characteristic behavior of ESH if the synthetic 
helices are replaced with real ones made of A, T, G, and C bases. Here, in this sub-section, we essentially focus on that. We consider 
two different cases: one where the bases A and G are arranged randomly, and another where the bases T and C are placed in a random 
sequence. The helical parameters ($R$, $\Delta h$, $\Delta \phi$ and $l_c$) associated with the SRH and LRH helices remain the same 
as those mentioned in Table~\ref{tab1}.
In these helices, we select the site energies associated with these four bases and the NNH integrals among them from 
the data sets available in the literature (see Refs.~\cite{sh6,zil}, and the references therein), and here we specify them as follows. 
The site energies of four different bases are: $\epsilon_A=8.631\,$eV, $\epsilon_T=9.464\,$eV, $\epsilon_G=8.177\,$eV, 
$\epsilon_C=9.722\,$eV.
\begin{figure}[ht]
\vskip 0.5cm
\noindent
{\centering \resizebox*{8cm}{8cm}{\includegraphics{Cv_T_ATGC.eps}}\par}
\caption{(Color online). Electronic specific heat as a function of temperature for the SRH and LRH helices composed of A, T, G and C bases. 
Two cases are considered: (i) helices with A-G bases and (ii) helices with T-C bases. Bases in each helix are arranged randomly, and 
averaging over 100 random configurations is performed. Two different field scenarios are 
taken into account, with the phase factor $\beta$ is set to zero. The system size and the filling factor are the same with those in 
Fig.~\ref{cvtvg}.}
\label{atgc}
\end{figure}
In the helix with A and G bases, the NNH strength between the G and G bases is $0.053\,$eV, while between the bases A and A
it is $-0.004\,$eV. Between the bases A and G, the NNH strength is $-0.077\,$eV in one direction and $-0.01\,$eV in the other direction 
(hoppings are direction-dependent). For the helix with T and C bases, the NNH strength is $0.18\,$eV between the T and T bases, and it 
is $0.022\,$eV between the C and C bases. For the unlike bases, i.e., between T and C, the NNH strength in one direction is $-0.028\,$eV
and it is $-0.055\,$eV in the other direction.
With these parameter setting, the characteristics of ESH with temperature for the SRH and LRH helices are shown in Fig.~\ref{atgc}. 
Two different field cases are taken into account. It is important to note that, unlike synthetic helices with $\epsilon_i=0$ for all 
sites, the helices here are originally disordered even when the transverse electric field is absent. This is because the site energies 
of the bases are different and the bases are arranged randomly. In presence of the electric field, an
additional modification is site energies occurs following Eq.~\ref{equ4}. A closer inspection of the spectra given in Fig.~\ref{atgc} 
reveals that the nature of $C_v$-$T$ curves is quite similar to those already discussed for the custom-designed helices. As 
$V_g$ increases, the difference between the two helices decreases, corroborating our earlier analysis.
It is worth mentioning that different sequences of A, T, G, and C bases in the helices almost lead to similar behavior of ESH, as confirmed
through our detailed numerical analysis. The main difference between a real helical system with A, T, G, and C bases and our chosen synthetic helices with a transverse electric field is that the latter can be utilized as ordered or disordered helices, depending on whether they are subjected to an electric field. Moreover, the response can be selectively tuned by externally regulating the field strength.
On the other hand, since real helices are originally disordered, tuning of ESH by means of an external electric field is also possible, 
but the response will be weaker than that of synthetic helices.

\subsection{Specific role of $t_1$ on electronic specific heat}

All the results of the custom-designed helices discussed above are computed considering the nearest-neighbor hopping strength 
$t_1=0.02\,$eV, so that $t_1$ becomes comparable to that in real helical molecules. In this situation situation, the electronic specific
heat peaks around the room temperature. Now an obvious question arises: how does the peak change if $t_1$ is modified? To illustrate it, 
in Fig.~\ref{cvtkb}, we present the results of electronic specific heat for both the SRH and LRH helices at three typical values of $t_1$.
\begin{figure}[ht]
\vskip 0.5cm
\noindent
{\centering \resizebox*{8cm}{4.25cm}{\includegraphics{Cv_T_t1.eps}}\par}
\caption{(Color online). Electronic specific heat as a function of temperature for the SRH and LRH helices in the 
field-free case, at three distinct values of $t_1$ where the blue, magenta and green curves are for $0.025$, $0.05$, and $0.1$, 
respectively.}
\label{cvtkb}
\end{figure}
For both helices, it is observed that the temperature at which the ESH reaches its peak value increases as the strength of $t_1$ is 
enhanced. Understanding this is straightforward, as the average energy correlates directly with all the physical parameters of the system, 
in addition to $K_B$. Given that $K_B$ scales at $10^{-5}$, it is recommended to reduce the strength of NNH in custom-designed helices 
to fully grasp the ESH across the accessible temperature range.

\section{Closing remarks}

To conclude, this study represents a pioneering investigation into the thermal signature of custom-designed and real 
helical systems, with a focus on the 
examination of electronic specific heat. While significant research has delved into charge and spin-dependent transport within various 
helical systems, the discourse on their thermal properties remains notably constrained. By studying the electronic specific heat, we 
gain valuable insights into how a substance manages the transfer and absorption of energy. The understanding derived from ESH not only
contributes to fundamental knowledge but also has practical implications. 

Depending on the physical parameters, two distinct types of helical systems are considered. Despite their structural resemblance, one 
helical system allows electrons to hop across a few neighboring lattice sites, while in the other system, all possible hoppings are 
permitted. Every helix is exposed to an external electric field oriented perpendicular to its axis. Simulating the helices within a 
tight-binding framework, we calculate the ESH by taking the first-order derivative of the average energy. The average 
energy is determined by fixing the number of electrons, with the electrochemical potential computed self-consistently. The results 
are analyzed for both field-free and finite-field conditions under spinless and spinful scenarios. A significant utility 
of our findings is also discussed. The key and novel discoveries of our work are summarized as follows.\\
$\bullet$ In the low-temperature regime, the SRH helix demonstrates a higher ESH compared to the LRH helix. Conversely, this trend 
reverses in the high-temperature regime. \\ 
$\bullet$ The disparity in ESHs between SRH and LRH helices gradually diminishes with increasing field strength. \\
$\bullet$ The thermal response of the helices can be selectively modulated by the electric field. \\
$\bullet$ The interaction between the range of electron hopping and the temperature regime plays a significant role in determining 
the thermal response. \\
$\bullet$ The results are not highly sensitive to the chirality of the system. Both right-handed and left-handed configurations can 
be employed almost equally. \\
$\bullet$ Finally, through the analysis of ESH, it becomes possible to estimate defective helices and identify the locations of defect 
sites. This could be particularly intriguing for diagnosing faulty cells and various diseases.

Our analysis can be employed to explore thermal responses in various types of these intriguing helices in the presence of higher-order 
electron hopping.

\section*{ACKNOWLEDGMENTS}

SS is thankful to DST-SERB, India (File number: PDF/2023/000319) for providing her research fellowship.
DL acknowledges partial financial support from Centers of Excellence with BASAL/ANID financing, AFB220001, CEDENNA.
We would like to thank all the reviewers for their constructive criticisms, valuable comments, and suggestions, 
which have greatly improved the quality of our work.

\end{document}